\documentclass[11pt]{article}
\usepackage[english]{babel}
\usepackage[utf8]{inputenc} 
\usepackage{geometry} 
\usepackage{color}
\usepackage{graphicx} 
\usepackage{amssymb}
\usepackage{amsmath}
\newcommand{\be}{ \begin{equation}}
\newcommand{\ee}{\end{equation} }

\fussy                 

\title{On the rotational dynamics of the Rattleback }
\author{Lasse Franti  \\
University of Helsinki \\
Department of Physics\\
and\\
Helsinki Institute of Physics\\
Gustav Hällströmin katu 2a \\
00560 Helsinki
}
\date{} 

\begin{document}   

\maketitle


The Rattleback is a very popular science toy shown to students all over the world to demonstrate the non-triviality of rotational motion. When spun on a horizontal table, this boat-shaped object behaves in a peculiar way. Although the object appears symmetric, the dynamics of its motion seem very asymmetric. When spun in the preferred direction, it spins smoothly, whereas in the other direction it starts to oscillate wildly. The oscillation soon dies out and the rattleback starts to spin in the preferred way.
We will  construct and go through an analytical model capable of explaining this behaviour in a simple and intelligible way. Although we aim at a semi-pedagogical treatise, we will study the details only when they are necessary to understand the calculation.
After presenting the calculations we will discuss the physical validity of our assumptions and take a look at more sophisticated models requiring numerical analysis.  We will then improve our model by assuming a simple friction force.
\newline


\section{Background}

For centuries the mysterious properties of the rattleback were used to predict the future and bring messages from other worlds. The alternative name "Celt" is thought to come from druids, or Celtic priests, who seem to have used the rattleback in their rites. This was probably a very good way to convince your followers of your powers, as the surprising behaviour looks quite magical even in our times. The priest was also able to adjust the predictions according to his needs by spinning the rock in the right direction. The peculiar motion of the object is still sometimes associated with all kinds of strange fields and ether forces. More conventional wrong explanations including the Coriolis force and the tennis racket theorem are also common, but fail to describe the situation correctly.

The first mathematically satisfactory analysis was written by Sir Gilbert Walker. In his paper "On a dynamical top" (1896) he assumed the celt to roll without slipping and neglected all dissipative forces. Walker also examined the motion assuming  purely oscillatory initial values. Walker's paper was not available to the author, but other sources (such as Garcia\&Hubbard) mention the analysis to be quite similar to the modern treatises with the same assumptions. Walker's paper is criticized for being quite difficult to follow, but his analysis has proved to be very resilient.
After the seminal paper of Walker, some celt-related papers were published in mostly educational publications. Interest in celt dynamics began to grow in the seventies and the 1980's became the golden age of celt dynamics. Many of the most important papers on the rattleback were published during a relatively short time.
Garcia\&Hubbard trace this sudden increase in interest to the appearance of J. Walker's popular article in Scientific American. These articles  concentrated mainly on numerical simulations of different celt models with dissipative forces and their significance in the dynamics.

 In 1986 Sir Hermann Bondi published  his classic paper "The Rigid Body Dynamics of Unidirectional Spin", in which he proposed an improved analytical model for the rattleback. Bondi's model forms the basis of most analytical treatises, although some of the credit should probably be given to G.T. Walker.  Bondi mentions the paper of Walker to be the only treatise he has been able to find, despite several papers  were published in the years between these two analyses. Bondi does not properly investigate the validity of his assumptions and the motivation of the spin reversal is not very rigorous. These are better addressed by Garcia\&Hubbard, who study the validity of the assumptions numerically and propose a better model for the spin reversal. Some other papers have also used  numerics to investigate the physical requirements of the conservative model. We will take a look at these investigations after our main treatise.

Some alternative proposals are quite unrealistic as such:  Caughey's celt model mentioned by Blackowiak\&al and Garcia\&Hubbard combined a viscous drag with springs parallel to the coordinate axes and, according to Caughey,  could "capture the essential features of the rattleback". However, Caughey's model could describe the motion perfectly and still fail to explain the behaviour of the celt,  as the system in question simply is not a celt with any reasonable assumptions.
 
Modern research in celt dynamics has become quite theoretical, as can be seen from the paper by Dullin\&Tsygvintsev whose treatise could well be regarded as mathematical physics.  Borisov\&Mamaev have also used modern techniques in their paper "Strange attractors in rattleback dynamics". With certain energies, the celt is shown to be an example of classical chaos in a conservative system. Despite its title, the paper also investigates the overall behaviour of the celt by using symmetries and phase mappings. The variables used in this paper are not very easy to link with the physical situation, which makes the interpretation of the various diagrams quite difficult.

One of the biggest strengths of the model presented here is the choice of simple variables. We will follow the contact point between the celt and the table, which can be expressed in the coordinates fixed to the celt. This choice is much more convenient than the usual Euler angles(e.g. Markeev), not to mention the Andoyaer-Deprit -coordinates of Borisov and Mamaev.
The fact that these kind of papers are still being published clearly shows that the classical mechanics of realistic systems is still far from being completely sorted out.

\section{Equations of motion and unidirectionality}

We will now present a simple model for the celt behaviour. We shall begin by looking at the stability of rotation and continue with an improved analysis of the spin reversal.
To explain the observed unidirectionality, we will present an adaptation of Bondi's classic treatise which forms the basis of most studies of celt motion. We shall try to add some comments to the equation-dominated presentation of the original article and correct its typographical errors.

Let us first introduce the notation:
\newline

\noindent
\begin{math}
\mathbf{s}=\textrm{position of the center of mass} \\
\frac{d \textbf{s}}{dt}=\mathbf{v}\\
\mathbf{r}=\textrm{vector from the center of mass to the point of contact}\\
\mathbf{F}=\textrm{force exerted to the celt by the table}\\
\mathbf{\omega}=\textrm{angular velocity of the celt}\\
\mathbf{h}=\textrm{angular momentum of the celt}\\
M=\textrm{mass of the celt}\\
\mathbf{u}=\textrm{upward pointing unit vector perpendicular to the table}
\end{math}
\newline
Using these we can write the equations of motion
\begin{gather}
M \frac{d \mathbf{v}}{d t}=\mathbf{F}-Mg \mathbf{u} \\
\mathbf{v}+\mathbf{\omega} \times \mathbf{r}=0 \label{veq} \\
\frac{d \mathbf{h}}{dt}=\mathbf{r}\times \mathbf{F} \label{perustorq}
\end{gather}
which give
\be
\frac{d \mathbf{h}}{dt}=M \mathbf{r}\times[\frac{d \mathbf{v}}{dt}+g \mathbf{u}] \label{hder}.
\ee
Taking a scalar product with \textbf{u} and swapping the operations then leads to
\be
\frac{d}{dt}(\mathbf{u} \cdot \mathbf{h})=M(\mathbf{u}\times \mathbf{r})\cdot \frac{d \mathbf{v}}{dt}.
\ee
At this point Bondi discusses the interpretation of this equation in ordinary rotational motion. We shall move straight into the rattleback case and define the form of the rattleback accordingly.
We shall thus define an object whose axes of inertia differ from the symmetry axes. Bondi arranges this by fixing the coordinates to the center of mass and choosing the axes of inertia to be the coordinate axes with the z-axis pointing downward at rest position. The moments of inertia are A, B, and C with $A>B$.
The form of the object is then defined to be 
\be
z=a\left[1-\frac{1}{2}p\left(\frac{x}{a}\right)^2-q\frac{xy}{a^2} - \frac{1}{2}s\left(\frac{y}{a}\right)^2  \right] \label{zeq}
\ee
where a is the equilibrium distance form the center of mass to the point of contact. This form can also be taken to represent the form only to second order near the equilibrium, in which case the object does not have to be an ellipsoid. First order terms are absent, as the point of contact is an extremum.

Physical validity limits the choice of the parameter values. The surface has to be concave, which requires
\be
p>0, s>0 \textrm{ and }ps>q^2.
\ee
This corresponds to a maximum of z at the origin.

We must also require the equilibrium position to be stable for the oscillation to take  place. This means that the radii of curvature have to be larger than the equilibrium distance a, which is satisfied if
\be
1>p, 1>s \textrm{ and } (1-p)(1-s)>q^2. \label{stabcond}
\ee

If the second order form is only an approximation, there has to be a safety margin to ensure stability during the peak amplitude. For typical celts both requirements are obviously satisfied. We shall thus assume the parameters to be physical and reasonable.

The celt will be analyzed by using a system of coordinates fixed to the celt itself. The contact point is now given by the vector $\textbf{r}=(x,y,z)$ where z is obtained from (\ref{zeq}).
The unit normal vector of the table can be obtained from the definition, as the unit normal of the celt surface at the point of contact has to coincide with the unit normal of the table. Taking the gradient of (\ref{zeq}) results in
\be
\mathbf{u}=-w\left(\frac{px+qy}{a}, \frac{qx+sy}{a},1\right), \label{ueq}
\ee
where $w$  the normalization factor given by
\be
w^{-1}=\sqrt{\left(\frac{px+qy}{a}\right)^2+\left(\frac{qx+sy}{a}\right)^2+1}. \label{normeq}
\ee
The normal vector and the third coordinate axis point at opposite directions at equilibrium, as expected.

\includegraphics[width=0.9\textwidth]{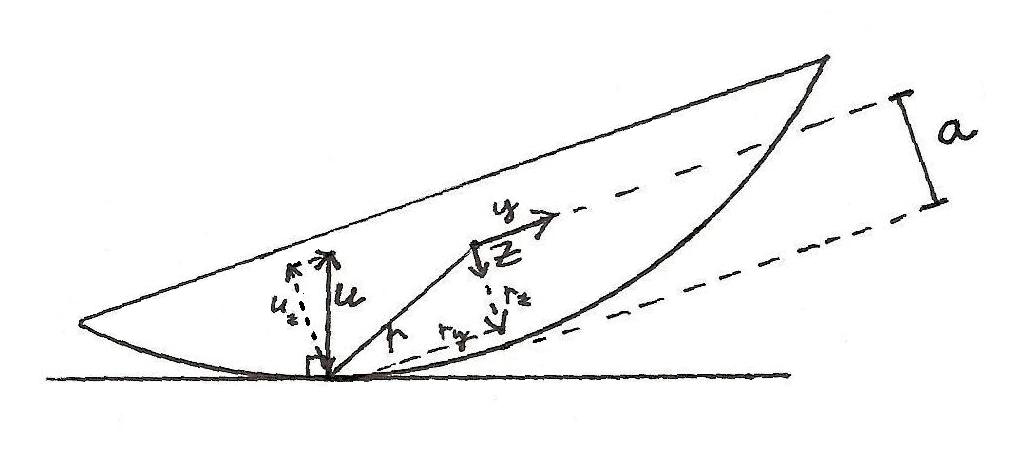}

 As \textbf{u} is constant in inertial coordinates, we can write
\be
0=\frac{d \mathbf{u}}{dt}=\dot {\mathbf{u}}+ \mathbf{\omega} \times \mathbf{u}.
\ee
where the dot thus stands for time derivative in the system defined by the inertial axes.

Taking a cross product with \textbf{u} and using the expansion formula for triple product results in the expression
\be
\mathbf{\omega}=\dot{\mathbf{u}}\times \mathbf{u}+n \mathbf{u},
\ee
where n is the "spin" or the vertical component of the angular velocity.  
We can expand the components of the angular momentum assuming x and y to be small enough for higher order terms to be dropped. The spin is assumed to be reasonable  but it is not explicitly assumed to be small. 

The z-component dominates if it is present, which enables us to write

\be
\begin{gathered}
\omega_1=\frac{1} {a} \left[q \dot x+s \dot y-n(px+qy)\right]  \\
\omega_2=\frac{1}{a} [-p \dot x-q \dot y-n(qx+sy)]  \\
\omega_3=\frac{1}{a^2} (ps-q^2)(\dot x y-\dot y x)-n[1-\frac{1}{2} (\frac{px+qy}{a})^2 -\frac{1}{2} (\frac{qx+sy}{a})^2]. \label{omegaeq}
\end{gathered}
\ee
  
The coordinate axes are the axes of inertia, which allows us to write 
\be
h_1=A \omega_1, h_2=B \omega_2, \textrm{ and } h_3=C \omega_3.
\ee

From (\ref{veq}) we obtain at first order
\be
\begin{gathered}
v_1= [p \dot x+q \dot y+n(qx-(1-s)y)] \\
v_2= [q \dot x+s \dot y+n((1-p)x+qy)] \\
v_3= \textrm{second order}.
\end{gathered}
\ee
Here all the terms arise from the zeroth order terms in z and the third component of the angular momentum. By inserting these expressions to (\ref{hder}) and taking into account that $d \mathbf{P}/dt=\dot{\mathbf{P}}+\mathbf{\omega}\times \mathbf{P}$ for all vectors $\mathbf{P}$, we can write the first component of the equation as
\be
\begin{gathered}
A \dot \omega_1 -(B-C)  \omega_2 \omega_3 \\ =\frac{A}{a}[q  \ddot{x}+s \ddot{y}-n(p \dot{x}+ q \dot{y})- \dot{n}(px+qy)]- \frac{n}{a}(B-C)[p \dot x+q \dot{y}+n(qx+sy)] \\ = -Ma \lbrace q \ddot{x}+s \ddot{y}+n(1-2 p) \dot{x}-2 n q \dot{y}+ \dot{n}[(1-p) x-qy]-n^2[qx-(1-s)y] \rbrace \\ +Mg[qx-(1-s)y] \label{ekakompo}
\end{gathered}
\ee
Here the first equality comes from expanding the expression to first order. The first term is the derivative of angular momentum and the second term comes again purely from the the third component of angular velocity. Second equality is the equality of equation (\ref{hder}). When expanding the expressions, it is convenient to remember at each step that we are only keeping terms up to first order. 

Similarly we obtain the second and third components of (\ref{hder})
\be
\begin{gathered}
B \dot \omega_2-(C-A) \omega_3 \omega_1 \\ =-\frac{B}{a}[p \ddot x+q \ddot y+n(q \dot x+s \dot y)+ \dot n (qx+sy)]+\frac{n}{a}(C-A) [q \dot x+s \dot y-n(px+qy)] \\= Ma \lbrace p \ddot x+q \ddot y+n[2q \dot x-(1-2s) \dot y]+\dot n[qx-(1-s)y]+n_2[(1-p)x-qy] \rbrace \\+Mg[x(1-p)-qy] \label{tokakompo}
\end{gathered}
\ee
and
\be
\begin{gathered}
C \dot \omega_3-(A-B) \omega_1 \omega_2 =-C \dot n[1-\frac{1}{2} (\frac{px+qy}{a})^2-\frac{1}{2}(\frac{qx+sy}{a})^2] \\+ {\frac{1}{a^2}} Cn[(px+qy)(p \dot x+q \dot y)+(qx+sy)(q \dot x+s \dot y)]  + \frac{1}{a^2} C(ps-q^2)(\ddot x y- x \ddot y) \\ = Mx \lbrace q \ddot x+s \ddot y+\dot n[(1-p)x-qy]  \rbrace -My  \lbrace p \ddot x+q \ddot y+ \dot n[qx-(1-s)y] \rbrace \\ +Mn \lbrace x[(1-2p) \dot x-2q \dot y]-y[2q \dot x-(1-2s) \dot y]-xn[qx-(1-s)y]-yn[(1-p)x-qy] \rbrace \\-M \frac{g}{a} [q(x^2-y^2)-(p-s)xy]. \label{spinder}
\end{gathered}
\ee
 The coefficient of $\dot n$ is of zeroth order, whereas the rest of the equation is of second order. This means that $\dot n$ itself must be of second order, which can be used to simplify the other two equations. We will further simplify them by adopting the following shorthand notation
\be
A+Ma^2=\alpha Ma^2, B+Ma^2=\beta Ma^2, C=\gamma Ma^2. 
\ee
 These allow us to write the equations (\ref{ekakompo}) and (\ref{tokakompo}) in the form
\be
\alpha(q \ddot x+s \ddot y)-(\alpha+\beta-\gamma)n(p \dot x+q \dot y)-(\beta-\gamma)n^2(qx+sy)+n \dot x+n^2 y=\frac{g}{a} [qx-(1-s)y] \label{ekakomposimpli}
\ee
and
\be
\beta(p \ddot x+q \ddot y)+(\alpha+\beta-\gamma)n(q \dot x+ s\dot y)-(\alpha -\gamma)n^2(px+qy)-n \dot y+n^2 x=\frac{g}{a} [qy-(1-p)x] \label{tokakomposimpli}
\ee
 We will need equation (\ref{spinder}) only with small values of n, when it can be written as
\be
a^2 \gamma \dot n=\gamma(ps-q^2)(\ddot x y-\ddot y x)+[q (y \ddot y-x \ddot x)+py \ddot x-s \ddot y x]+\frac{g}{a} [-(p-s)xy+q(x^2-y^2)] \label{spindersimpli}
\ee
Now we can take a look at the solutions of the simplified equations of motion (\ref{ekakomposimpli}) and(\ref{tokakomposimpli}). As we are examining an oscillation around the equilibrium point, it is natural to assume a solution of the form $C e^{\sigma t}$. By inserting this solution to the equations and cancelling the exponent, we arrive at a system whose characteristic equation is
\be
\begin{gathered}
(\sigma^2+n^2) [ \alpha \beta(ps-q^2) \sigma^2 +(\alpha - \beta)qn \sigma] \\ + (\sigma^2+ n^2) \lbrace n^2[1-p(\alpha-\gamma)-s(\beta-\gamma)+(ps-q^2)(\alpha-\gamma)(\beta-\gamma) ]   +\frac{g}{a} [\alpha s+ \beta p-(ps-q^2)(\alpha+\beta)] \rbrace     \\  + \frac{n^2 g}{a} [ 2-(1+ \alpha+ \beta-\gamma)(p+s)+2(\alpha+\beta-\gamma)(ps-q^2)]+(\frac{g}{a})^2 [(1-p)(1-s)-q^2]=0
\end{gathered}
\ee

This equation determines whether the system has nonzero solutions. The characteristic is of fourth order, which would allow for a explicit solution by using the Cardano formulae. Fortunately, we do not have to solve this equation to understand the behaviour of the celt as the important features can be seen from the equation itself. 
The roots of the equation depend on the direction of spin in a straightforward way, namely:
\be
\begin{gathered}
\textrm{If } \sigma \textrm{ is the solution of the characteristic equation for some value of n,} \\ \textrm{ the solution for the value -n is -}\sigma. \nonumber
\end{gathered}
\ee

This is easy to see, as the variables are always squared except in one term in which they are both present. It is worth noticing that our model is fully conservative. The diminishing amplitude is not caused by dissipation, but is an indication of energy flow between the oscillation and the rotation.

The behaviour depends on the overall distribution of signs. The main classes are the following:
\newline
(a) The real parts of the roots all have the same same sign on some interval of spin values. This makes one of the spin directions stable and the other unstable.
\newline
(b) The real parts of the roots have different signs on some interval of spin values, which leads into unstable motion in both directions. In both directions energy flows into the oscillations corresponding to positive real parts. Some positive roots may have much smaller absolute values than their negative counterparts, which would result in the rotation momentarily gaining energy from the oscillation.

We are mostly interested in case (a), which essentially the explains the behaviour of commercial rattlebacks. The direction of energy flow depends on the direction of spin, which causes the observed difference in stability.

According to Bondi's linearized model, unidirectionality is observed when a body of suitable form has a stable equilibrium and unequal principal moments of inertia with axes of inertia differing from axes of symmetry.  If these conditions are not met, the coefficient $(\alpha-\beta)q$  multiplying the  important cross term will vanish.
In fact, unidirectionality requires the radii of curvature to be different, too. All the requirements are beautifully expressed in Bondi's further studies and in the analysis of torque by Garcia\&Hubbard. We will study both of these later on in this paper.

The difference in the axis alignment is caused by the coefficient q, but the object can also be unsymmetrized by adding weights to a symmetric body. Most experiments have used commercial rattlebacks, whose properties have to be estimated from the dynamics itself. Taking a symmetric body and adding weights into it would provide us with an object with known inertia tensor and shape. This would enable us to make more accurate comparisons between the simulations and observed behaviour. The author has made a few celts this way and they work surprisingly well. The post-reversal spin rates are often left smaller than in commercial celts but the unidirectional behaviour is quite easy to observe even with relatively rustic models.

Bondi then analyzes the relation between the sign distribution and celt shape. This results in a diagram classifying celts into types 0,I and II, although the zero region was not given a special name in the original article.
 In this classification 0-type corresponds to instability in both directions and I-type to unidirectionality. According to Garcia\&Hubbard, type II behaves similarly to the unidirectional type, but the axis of rotation differs from the vertical equilibrium position.
The behaviour of the celt thus depends on its position in the chart. The parameters of the celt affect its stability and by fixing some of them we can find threshold values for the rest. In addition to the linearized model, this belongs to the classic material of Bondi's paper. Some aspects of this analysis are presented in the appendix.

After classifying the various celt types, Bondi considers the spin reversal. Rotation in the unstable direction comes to a halt while an oscillation remains. Equation (\ref{spindersimpli}) shows, that the spin cannot remain zero, but the celt starts to spin again. According to Bondi, it is very implausible that the object would start spinning in the same direction after having just stopped due to the instability. The object would thus start spinning in the stable direction and the negative real parts would pump the energy into the spin. In principle, Bondi's model thus explains the observed spin bias.

\section{Torque and the spin reversal}

The explanation given in Bondi's final conclusions is quite credible but as an essential feature of celt dynamics the spin reversal deserves to be studied more thoroughly. The reversal is investigated in detail by the mostly numerical paper by Garcia\&Hubbard, in which the authors present in a rather concise way a model based on average torque. We will now go through their reasoning and comment on its details after the calculation.

Similarly to the previous model, Garcia\&Hubbard neglect dissipation and assume the celt to be rolling without slipping. The initial situation is a slowly rotating celt with energy stored in the oscillation.

We are interested in the vertical torque which is obviously equal to the vertical component of the derivative of angular momentum:
\be
\tau_v= \mathbf u \cdot d \mathbf h/dt=M \mathbf u \cdot (\mathbf r \times d \mathbf v/dt)= M(\mathbf u \times \mathbf r) \cdot d \mathbf v /dt \label{torq}
\ee
Integration by parts gives 
\be
\frac{1}{ M} \int_{T} \mathbf u \cdot d \mathbf h=\frac{T}{M} \hat   \tau_v= [(\mathbf u \times \mathbf r) \cdot \mathbf v ]_T - \int_T \mathbf v \cdot (\mathbf u \times d \mathbf r+ d \mathbf u \times \mathbf r)
\ee
As the motion is approximately periodic during the interval T, we obtain

\be
[(\mathbf u \times \mathbf r) \cdot \mathbf v ]_T=0
\ee
By writing $\mathbf{v}=\mathbf{r}\times \mathbf{\omega}$ and $\mathbf{r}=\mathbf{Ru}$ we can express this as

\be
\frac{T}{M} \tau_v=- \int_T (( \mathbf R \mathbf u) \times \mathbf {\omega}) \cdot (\mathbf u \times( \mathbf R d \mathbf u) + d \mathbf u \times ( \mathbf R \mathbf u)),
\ee
where the matrix \textbf{R} has entries
\be
\begin{gathered}
r_{11} /a=- \frac{1}{2}[(1+\psi)/\theta+(1-\psi)/\phi]=-\cos^2 \xi/\theta-\sin^2 \xi/\phi \\ r_{12}/a =r_{21} /a=-\frac{1}{2}(1/\phi-1/\theta)(1-\psi^2)^\frac{1}{2}=-(1/\phi-1/\theta) \cos \xi \sin  \xi \\ r_{22} /a=- \frac{1}{2}[(1-\psi)/\theta+(1+\psi)/\phi]=-\sin^2 \xi/\theta-\cos^2 \xi/\phi \\
r_{13}=r_{23}=r_{31} =r_{32} =0 \\ r_{33} /a =-1
\end{gathered}
\ee
The original paper lacks a minus sign in the first equation. This is probably just a printing error, as all the results are consistent with the correct form.

Garcia and Hubbard define the curvature parameters as follows: 

 \noindent
\begin{math}
\theta= \textrm{the ratio of a to the smaller radius of curvature} \\
\phi= \textrm{the ratio of a to the larger radius of curvature} \\
\psi=\cos^2 \xi-\sin^2 \xi,
\end{math} \newline
where $\xi$ is the small angle between the direction of the smaller radius of curvature and the axis of inertia deviating from it. 

Bondi uses the the same parameters but defines them in a completely different way.  The definitions are equivalent, as can be seen by diagonalizing the quadratic form and using the formula for curvature radius. The cosine of the skew angle can be computed from these results and the use of elementary trigonometry gives the result. Some of Bondi's results have to be reformulated to see the connection better. We did not present the classification of celt types explicitly but the equivalence of the definitions and the calculations involved are quite illuminating as such. Among other things, they explain the stability condition (\ref{stabcond}) postulated by Bondi.  Despite its algebraic nature, we thus present the analysis in the appendix.

Expanding the integral up to fourth order results in the expression

\be
\frac{T}{M} \tau_v=a \int_T [\omega_1 du_1(a+ r_{11})+\omega_1 du_2 r_{12 }+ \omega_2 du_1 r_{21}+ \omega_2 du_2 (a+r_{22})] \label{approtorq}
\ee
where the quantities involved are obtained from linearized equations.

The derivative of the state vector $[\omega_1 \phantom k \omega_2 \phantom k  u_1 \phantom k u_2]$ can be obtained from the vector itself by multiplying with the matrix

\be
\mathbf{A}=\left[ \begin{array}{cccc}
0 & 0 & d& e \\
0&0 & f & k \\
0 & 1 & 0&0 \\
-1&0&0&0
\end{array} \right],
\ee
where the coefficients are defined as

\be
d=-\frac{g r_{11}}{\alpha a^2}, \quad e=-\frac{g(a+r_{22})}{\alpha a^2}, \quad f=\frac{g(a+r_{11})}{\beta a^2}  \quad k=\frac{g r_{12}}{\beta a^2}
\ee
The eigenvalues of \textbf{A} are
\be
\lambda^2_{1,2}=-\frac{1}{2}\lbrace (e-f)\pm[(e-f)^2+4(ef-dk)]^\frac{1}{2}\rbrace
\ee
If the geometric axes coincide with the axes of inertia, we obtain the squared frequencies $e$ and $-f$. We thus define the two eigenvalues to be $\lambda^2_{1,2}=-e' , f'$. The initial conditions are obtained from the eigenvectors:
\be
\mathbf{V}=
\left[ \begin{array}{cccc}
-e(e'+f)+dk & -e(e'+f)+dk & f'd& f'd \\
-e'k&-e'k & f(f'+e)-dk &f(f'+e)-dk \\
i\sqrt{e'}k &- i\sqrt{e'}k & i\sqrt{-f'}(f'+e)&- i\sqrt{-f'}(f'+e) \\
-i\sqrt{e'}(e'+f)&i\sqrt{e'}(e'+f)&-i\sqrt{-f'}d& i\sqrt{-f'}d
\end{array} \right]
\ee
 
Initial conditions $\frac{1}{2}(\mathbf{v_1+v_2})$ and$\frac{1}{2}(\mathbf{v_3+v_4})$ give the torques
\be
\begin{gathered}
\hat \tau_{v^e} = \frac{1}{4} M a^2 \alpha (\frac{1}{\phi} -\frac{1}{\theta})(\frac{1}{\alpha} -\frac{1}{\beta})(1-\psi^2)^\frac{1}{2} e'^2 [e(e' +f)-dk] v^2_e \\
\hat \tau_{v^f} = \frac{1}{4} M a^2 \beta (\frac{1}{\phi} -\frac{1}{\theta})(\frac{1}{\alpha} -\frac{1}{\beta})(1-\psi^2)^\frac{1}{2} f'^2 [f(f' +e)-dk] v^2_f ,
\end{gathered}
\ee
where $\mathbf{v_{1,2,3,4}}$ are the column vectors of $\mathbf{V}$ and $v_f, v_e$ are normalization factors with dimension $\textrm{s}^3$.

\section{Commentary}

The first part of the calculation by Garcia\&Hubbard  is more or less a straightforward application of rigid body dynamics. Expression (\ref{torq}) has the same form in the  rotating frame of reference, which can be shown by using simple vector algebra. Matrix $\mathbf R$ can be derived from Bondi's equation (\ref{ueq}) by inverting the relation and using equations (\ref{curparbondi}) to transform the result to these variables. The calculation becomes easier, if one notices the relation $ps-q^2=\theta \phi$ .
The entries of the dynamical matrix $A$ can be derived from (\ref{perustorq}) by using the relation between $ r$ and $u$. The horizontal forces near the equilibrium point can be found by using the components of $u$, which results in the expression given in Garcia\&Hubbard.

After obtaining the various components, the calculation becomes rather mechanical. If the skew angle vanishes,  the entries of the matrix can be deduced by drawing a picture. The torques can be found quite easily and the result can be written down. In both methods one has to pay special attention to the signs, especially as the vertical axis points downwards.
The skew angle enters the expressions in $\mathbf{A}$ and $\mathbf{R}$ only because the coordinate axes differ from the curvature axes. In most celts this angle is quite small, which allows us to study the connections between the various quantities more easily by temporarily assuming $\xi=0$. This assumption is not possible in our actual treatise, as the asymmetry is necessary for the celt to work.  
 
The substitution term in the integral is assumed to vanish due to the periodicity. The motion is not absolutely periodic and the forces considered rather small. The motion is approximately periodic during one cycle, which makes the substitution negligible.
 
The linearized equations mentioned by Garcia\&Hubbard are ambiguous, as their paper does  not explicitly present clear candidates. One possibility is to use equations (\ref{ueq}),(\ref{normeq}), and (\ref{omegaeq}) assuming the spin to be small.  Taking terms up to second order into account one obtains the same form of equation (\ref{approtorq}).  It can be shown that the squared frequencies -f' and e' are positive real numbers. The choice of signs ensures that the primed frequencies take their unprimed values when the skew angle approaches zero. The eigenvectors are obtained from A and choosing a suitable linear combination results in a trigonometric form in both modes. According to the dynamical matrix the derivatives of $u$ are the angular frequencies, which can be verified from the eigenvectors. This connection comes handy in simplifying the expressions. The normalization factors fix the amplitude of the oscillations and correct the dimension of the initial values.
The diminishing amplitude of the oscillation is taken into account through the dynamics included in the linearized equations, which causes the final state to differ from the initial one. The averaging leads into integrating the squares of trigonometric functions over the period and a nonzero result.
Simplifying the result after the integration is quite tedious, but results in the expression obtained by Garcia\&Hubbard.
 We have presented this analysis mainly to verify Bondi's conclusions, which makes the technical details and the exact form of the result not so important. The calculation verifies that a suitably asymmetric body with two unequal principal moments of inertia and two differing radii of curvature oscillating on a horizontal table experiences a nonzero torque, which causes the spin reversal. It is very pleasing to note that exactly the same conditions are obtained by Bondi during the classification of celt types. The intermediate steps presented in the commentary probably differ somewhat from the original path taken by Garcia\&Hubbard. We could thus claim the conditions to be obtainable in three at least somewhat different ways.
The diminishing spin and increasing oscillation are caused by the torque, which remains nonzero even after the celt comes to a halt. This nonzero torque is responsible the spin reversal. In the stable direction energy flows back to the rotation, which can already be seen from the preceding model by Bondi.

\section{Validity of models}

With the assumptions made here, Bondi's linearized model seems to be a generally accepted explanation for the unidirectional stability of the celt.
On the other hand, the true nature of the spin reversal has been discussed more frequently and many numerical papers give a larger role to the dissipative forces ignored by Bondi.
Garcia\&Hubbard use numerical simulations to show the negligibility of aerodynamic drag in the motion of typical celts made of relatively dense materials.  Although Garcia\&Hubbard use a relatively coarse linearized model, we can rule out the effects of the airfoil form.
Another class of models allow the contact point to slide. Studying  these models is probably in order, since maintaining a non-sliding contact seems to require unrealistically high coefficients of friction, as shown in the paper by Garcia\&Hubbard. The effects of sliding friction are investigated in the papers by Magnus, which seem to be unavailable. Taking into account the sliding friction and sliding in all directions will naturally lead into considerable algebra. It is therefore not very surprising that  Lindberg\&Longman report the discussion in Magnus to be too heavy to offer any insight into the motion. Lindberg\&Longman's simulations with the assumptions made in Bondi showed unidirectionality, which confirms the physical validity of the model.

The coefficient of friction can be taken to be very large, if we model the point of contact to be a mathematical point. This would prevent the slipping without slowing the celt down. This is not necessary or even desirable, as realistic celts do slow down. The various celt models also seem to predict extra reversals, if the simulation is run for unrealistically long times. Although multiple reversals can be observed with some celts, the most usual commercial types seem to reverse only once before coming to a halt due to friction. In the paper by Lindberg\&Longman, the extra reversals take place at 125 seconds and 150 seconds, depending on the initial direction of spin.  If the graphs are restricted to realistic spin times,however, we see a beautiful reversal followed by rotation in the stable direction.
Adding a dissipative vertical torque without any other dynamical effects would thus result in a rather realistic model. This could be done by replacing the mathematical contact point with an area of contact, which is in fact a better description of a contact between two objects. The celts built by the author seem to work the best when the initial spin is rather small and the oscillatory motion demanding the largest coefficients of friction thus not so violent. On the other hand, the existence of a vertical torque caused by the friction is clearly visible. The modest oscillation left behind cannot overcome the friction and the celt is left non-rotating. Although multiple reversals do not occur in our linearized model, assuming an area of contact still makes the model more realistic.

We can conclude that our assumptions are reasonably physical and our treatise thus a satisfying explanation for the dynamics of the celt.  According to simulations, higher order calculations would result in quantitatively reasonable results. Taking higher orders into account would make the necessary analytical calculations rather tedious.
\section{Conclusion}
By combining the classic Bondi model with a more accurate investigation into the asymmetric torque, we obtain an analytic model capable of explaining the behaviour of the rattleback without numerical calculations. The presentation given here provides a better opportunity to understand the dynamics of the rattleback.
\section*{Acknowledgements }
The author acknowledges support from Helsinki Institute of Physics and University of Helsinki Department of Physics. During the preparation of the English version the author has received support through a Finnish Academy of Science and Letters grant.

\section*{On the curvature parameters}

The curvature parameters $\theta$, $\phi$ and $\psi$ originate from Bondi's studies, where their role is to simplify the classification of celt types. This is done by studying the roots of the characteristic equation. Bondi defines the parameters as follows:
\be
\begin{gathered}
\theta+\phi=p+s \\ \theta \phi=ps-q^2 \\ \psi(\theta-\phi)=p-s 
\end{gathered}
\ee
According to this definition the parameters are independent of the sign of q or the direction of the skew angle. By choosing the sign of q  we simultaneously fix the direction of the skew angle and the stable direction of spin.

Choosing q to be positive and solving for the original parameters gives

\be
\begin{gathered}
   p= \frac{1}{2} (\theta+\phi)+ \frac{1}{2} \psi (\theta-\phi)  \\ q= \frac{1}{2} (\theta-\phi)(1-\psi^2)^\frac{1}{2}  \\ s= \frac{1}{2} (\theta+\phi)- \frac{1}{2} \psi (\theta-\phi). \label{curparbondi}
\end{gathered}
\ee
By defining $\rho=\sigma/n$ and $\Omega=g/an^2$ we can write the characteristic equation as
\be
(\rho^2 +1)(\rho^2 + \chi \rho+\kappa+\Omega \lambda)+\Omega \mu +\Omega^2 \nu=0,
\ee
where Bondi defines the new auxiliary parameters as follows

\be
\begin{gathered}
\alpha \beta \theta \phi \chi= \frac{1}{2} (\alpha-\beta)(\theta-\phi)(1-\psi^2)^{\frac{1}{2}} \\ \alpha \beta \theta \phi \kappa= 1-\frac{1}{2} (\alpha+\beta-2 \gamma)(\theta+\phi)+(\alpha-\gamma)(\beta-\gamma) \theta \phi- \frac{1}{2} (\alpha-\beta)(\theta-\phi) \psi \\ \alpha \beta \theta \phi \lambda= \frac{1}{2} (\alpha+\beta)(\theta+\phi-2 \theta \phi)- \frac{1}{2} (\alpha-\beta)(\theta-\phi) \psi \\ \alpha \beta \theta \phi \mu =2-(\theta+\phi)-(\alpha+\beta-\gamma)(\theta+\phi-2 \theta \phi) \\ \alpha \beta \theta \phi \nu=(1-\theta)(1-\phi) \label{charpara}
\end{gathered}
\ee
The real part of the root cannot change sign without vanishing at some point, which enables us to find the critical values separating different celt types.
In practice, we end up with conditions for the parameters $\mu$ and $\kappa$, which in turn restrict the curvature and inertial parameters through equations (\ref{charpara}). To draw the diagram mentioned earlier in the text, we need to fix some parameters. In the diagram taken from Bondi the values of the inertial parameters have been fixed and the division lines given with four different values of the skew angle.
Celts situated left to the line $\mu=0$ are unstable in both directions, whereas type I celts between the lines $\mu=0$ and $\kappa=0$ are unidirectional. The more complex type II celts lie right to the line $\kappa=0$. As mentioned earlier, the diagram shows four different cases with varying skew angle $\psi$.

\includegraphics[width=0.9\textwidth]{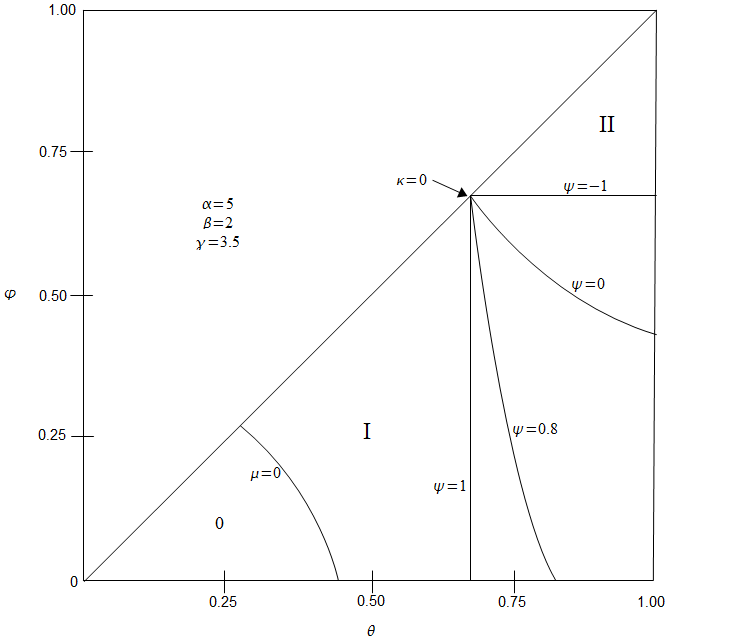}

The equations used to simplify the analysis are not very interesting as such, but they do reveal the conditions of unidirectionality to agree with the ones given by Garcia\&Hubbard. If these requirements are not satisfied,  $\chi=0 $ and the term causing the unidirectional behaviour will disappear. The coefficients $\alpha,\beta,\theta$, and $\phi$ cannot vanish for a physical object.

Although the conditions seem to agree, the curvature parameters involved here have been derived in a completely different way. The definition given here stems from their convenience in the analysis of roots, whereas in the paper by Garcia\&Hubbard these are associated with the geometry of the celt. The equations of Bondi are used in the analysis together with the ones derived in Garcia\&Hubbard, which requires the definitions to be equivalent. Establishing this relation is the main topic of this appendix. Although we are intrested in confirming the result, the calculation itself is quite illuminating as such.

To compare the definitions, we need to solve the curvature parameters from the definition given by Bondi:
\be
\begin{gathered}
\theta=\frac{1}{2} (p+s)+ \frac{1}{2} \sqrt{(p-s)^2+4q^2} \\ \phi=\frac{1}{2} (p+s)-\frac{1}{2} \sqrt {(p-s)^2+4 q^2} \\ \psi=\frac{p-s}{\sqrt{(p-s)^2+4 q^2}}
\end{gathered}
\ee
In the system defined by the axes of inertia, the object has an ellipsoidal form:
\be
z=a \left[  1-\frac{1}{2} p (\frac{x}{a})^2-q \frac{xy}{a^2}- \frac{1}{2} s (\frac{y}{a})^2 \right].
\ee
This can be expressed in a matrix form as
\be
 \left[ \begin{array}{cc}
\frac{1}{2}\frac{ p}{a} &   \frac{1}{2}\frac{q}{a}  \\
\frac{1}{2}\frac{q}{a}& \frac{1}{2}\frac{s}{a}   \\ 
\end{array} \right]
\ee
The eigenvalues needed in diagonalizing this matrix are
\be
\frac{\frac{1}{4} (p+s)+ \frac{1}{4} \sqrt{(p-s)^2+4q^2}}{a} \quad \textrm{and} \quad \frac{\frac{1}{4} (p+s)- \frac{1}{4} \sqrt{(p-s)^2+4q^2}}{a}.
\ee
Using these, we obtain the eigenvectors
\be
\left[ \begin{array}{cc}
\frac{-q}{\frac{1}{2} (p-s)- \frac{1}{2} \sqrt{(p-s)^2+4q^2}} \\
1
\end{array} \right]
\textrm{and}
\left[ \begin{array}{cc}
\frac{-q}{\frac{1}{2} (p-s)+ \frac{1}{2} \sqrt{(p-s)^2+4q^2}} \\
1
\end{array} \right]
\ee
In the system defined by the eigenvectors, the expression is diagonal. This corresponds to a symmetric ellipsoid i.e. the axes we have found are the axes of curvature. These are orthogonal as required. The other joins the corresponding axis of inertia smoothly as $q \rightarrow 0$, whereas the other approaches its counterpart when the lower component becomes negligible during normalization.

To find the angle between the axes, we need to normalize the eigenvectors. Let us choose the second vector and divide it by its length. From the scalar product of the normalized eigenvector and its counterpart $[0 \quad 1]^T$, we obtain 
\be
\cos \xi =\frac{1}{\sqrt{1+ \frac{q^2}{\left(\frac{1}{2} (p-s)+\frac{1}{2} \sqrt{(p-s)^2+4q^2}\right)^2}}}
\ee
Using the identity $\cos^2 \xi-\sin^2 \xi=2 \cos^2 \xi-1$ results in
\be
\cos^2 \xi-\sin^2 \xi=\frac{p-s}{\sqrt{(p-s)^2+4 q^2}}
\ee
The definitions of $\psi$ are thus equivalent.

 For verifying the two other parameters we need the formula for curvature radius
\be
R=\frac{\left[1+(\frac{dy}{dx})^2\right]^\frac{3}{2}}{\left| \frac{d^2 y}{dx^2}\right|} \label{radcur}
\ee
The curvature axes are the eigevectors and according to the general theory, the diagonal form contains the eigenvalues. By using formula (\ref{radcur}) to the diagonal form, we obtain the radii of curvature near the equilibrium to be
\be
R_1=\frac{a}{\frac{1}{2} (p+s)+ \frac{1}{2} \sqrt{(p-s)^2+4q^2}} \quad \textrm{and} \quad R_2=\frac{a}{\frac{1}{2} (p+s)- \frac{1}{2} \sqrt{(p-s)^2+4q^2}}.
\ee
The definition of Garcia\&Hubbard thus gives
\be
\theta=\frac{a}{R_1}=\frac{1}{2} (p+s)+ \frac{1}{2} \sqrt{(p-s)^2+4q^2} \phantom . \textrm{and} \phantom . \phi=\frac{a}{R_2} =\frac{1}{2} (p+s)- \frac{1}{2} \sqrt{(p-s)^2+4q^2}
\ee
in agreement with the definition by Bondi.

The stability condition (\ref{stabcond}) requiring the radius of curvature to be larger than the equilibrium height of the center of mass can now be understood more properly. This requirement is indeed satisfied if
\be
1>p, 1>s \textrm{ and } (1-p)(1-s)>q^2,
\ee
as previously stated.

Interestingly enough, it seems inevitable that Bondi must have done a similar calculation to establish the stability condition. Bondi has probably also solved his definition for the curvature parameters. From these two results the connection of the parameters to the celt shape is quite easy to see, but Bondi does not mention anything about their geometrical meaning. This is in retrospect easy to see, but fairly difficult to see directly from the equations without precognition, which may have caused it to remain unnoticed.

\section*{Mentioned papers}
Sir Gilbert Walker, 1896 \emph{Quarterly Journal of pure and Applied Mathematics} \textbf{28}:175-84 \newline
Magnus,K 1971 \emph{Theorie und Praxis der Ingenieurwissenschaften}  19-23     \newline
Magnus, K 1974 \emph{Zeitschrift für Angewandte Mathematik und Mechanik} \textbf{54}:54-5 \newline
Caughey, T.K. 1980 \emph{International Journal of Non-linear Mechanics} \textbf{15}:293-302

\end{document}